# Self-Healing Audio System

Shubham Sharma, Aditya Sridhar, Jai Prakash Krishnia
Harman International Industries, Incorporated
[shubham.sharma@harman.com, aditya.s@harman.com, jaiprakash.krishnia@harman.com]

*Abstract*— Installed sound applications typically involve a large number of audio processors, amplifiers and speaker systems spread across the venue. They could be spatially distributed at the venue across different rack rooms and floors. These systems are commissioned and configured by sound engineers using software application(s). This is essentially a one-time activity, following which, the audio systems run independently. Detection of faults and reconfiguration of any audio device(s) that fail(s) is a time-consuming operation. This disruption in the audio system can affect the entire audio chain and affect the usability of the venue in question. In this paper, we provide an overview of an audio system that detects the replacement for any faulty audio device(s) in the network and re-purposes the same to restore the configuration to last working point.

## I. INTRODUCTION

An audio system consists of several components, like digital signal processors, amplifiers, speaker systems and peripherals. These components may be distributed across the venue where the audio system is installed and connected by a common network backbone. The audio signals generated from the input devices like microphones are transmitted to digital signal processors that enhance them by applying audio effects (such as gain, equalization, filtering etc.). Following this, the signals are passed through audio power amplifiers which boost the signals to levels suitable for loudspeaker systems. This represents the typical audio signal flow in a venue (Figure 1).It is apparent that each component is vital for delivering the sound to the audience. Once a venue has been configured and commissioned by sound engineers, the maintenance of the audio systems is tasked to operators/technicians. It is expected that the audio system will work reliably and continuously. Failure of any of these components can be a critical issue, especially if a venue has gone 'live'.

On detecting any device failures, maintenance activity is initiated. The naive approach to overcome the failure is to connect and configure a new device with the same configuration of failed device. The replacement device has to be configured again to match the characteristics and the configuration of the failed device. The sound engineer would have to use the necessary tools, for e.g. software applications, to apply exactly the same configuration to the new device. This can contribute to increased turnover time for the maintenance activity.

There are other ways to solve this problem, for e.g. the audio devices could support redundancy [1] [2]. This approach would entail each audio device to have a backup that is connected to the same network as the primary device. On detecting the failure of the primary device, the secondary device can be activated and configured to take the place of the failed device in the system. But this solution is not scalable and requires significant capital investment in terms of having redundant devices.

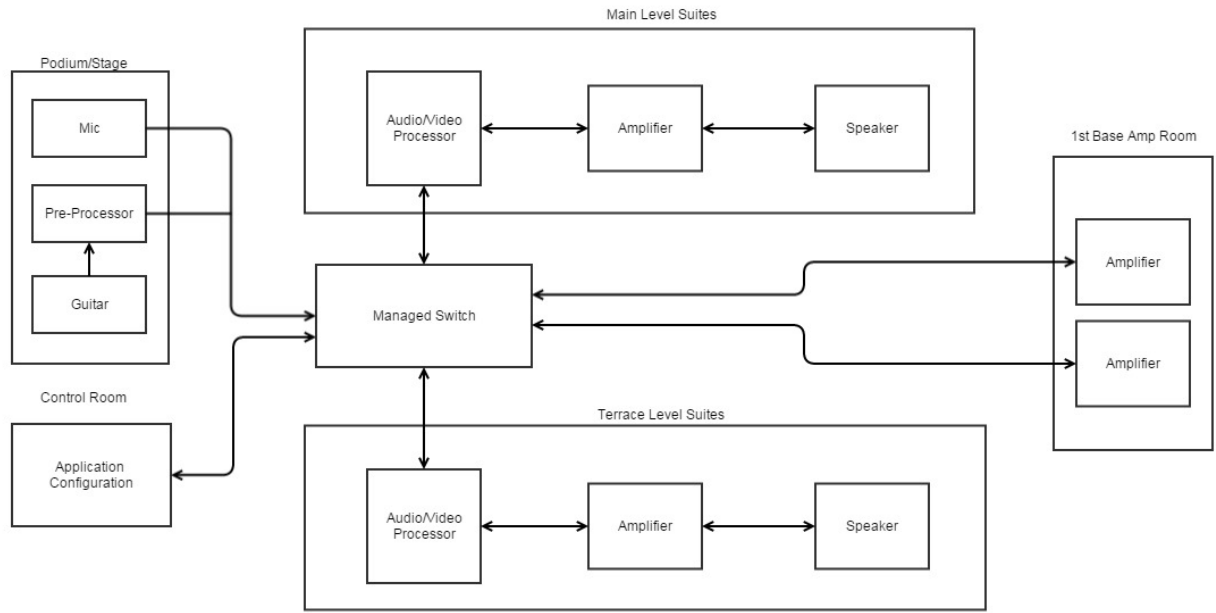

**Figure 1: Audio System**

The challenge is to provide a system that is able to automate the re-configuration of the devices with minimal human interaction and reduce the time required to restore the audio system to a fully operational state. The system that we propose here takes a different approach by removing the need to have a large number of redundant devices. This is scalable and only requires replacement when the device fails. These claims are supported by comparing performance of the system implemented with mentioned approach and naïve solution in which manual re-configuration is done.

Outline: In Section II, we describe our approach in brief. In Section III, implementation of system is discussed in detail. In Section IV, we discuss the experiments with our implemented solution with varying number of devices. Finally, we conclude in Section V.

## II. OVERVIEW OF THE APPROACH

Device detection on the network is achieved using a network monitoring protocol. As a result, we will know the status of devices connected to any port of a network switch. Detection of devices present on a network switch will be a continuous process.

Whenever, any device fails either by losing connection to network or device crash, we will replace the failed device with a fresh device. It is assumed that the user/operator has been notified of the failure of the device and has introduced the replacement devices into the network. The mechanisms for notifying the user about devices that have failed are not within the scope of this paper. All the audio devices that have been discovered newly on the network following the event of device failure are considered as candidates for replacing the failed devices. Replacement devices are searched and the best candidate is selected using Switch Port Detection Technique [3] and comparing device properties.

After the replacement candidates for the failed audio devices have been evaluated and the compatible replacements identified, the system proceeds to map the characteristics and the configuration of the failed devices onto the replacements.

It is important to note that the system has the ability to obtain the last working configuration of the audio devices. The files are transferred using a suitable file transfer mechanism such as File Transfer Protocol (FTP) [4]. The snapshot of the audio parameters is transferred using messaging conduits that all the audio devices are capable of communicating over.

The audio system is said to have healed a failed audio device when it has successfully completed the above-described processes with the replacement device. At this stage, the system has completely been restored to its pre-failure condition with minimal user interaction and without disrupting the functioning of the non-failed devices.

## III. IMPLEMENTATION

Simple Network Management Protocol (SNMP) [5] is used for network monitoring because this is supported by all network switches across brands i.e. HP, Cisco, Netgear.

The system will act as SNMP manager and Network Switches present on the network will act as SNMP agents. Network objects of the switch are stored in a database which is referred as Management Information Base (MIB) [6], which uses an Object Identifiers (OID) to keep track of these objects. Rules for naming objects, data types and how to encode information are defined using Structure of Management Information (SMI).

First, we need to do Network Switch identification and validate support for SNMP. To retrieve a list of devices connected to the ports of a network switch, its lookup table is accessed. Switch Lookup Table can be accessed by using a combination of object identifiers in SNMP [7].

Following are the steps for retrieving Switch Lookup Table:
1. Retrieve MAC Address Table, OID for this table is .1.3.6.1.4.1.9.9.46.1.3.1.1.2.
This table will map each port to connected devices as a list of MAC addresses on it.

2. Retrieve Port Number Table, OID for this table is .1.3.6.1.2.1.17.4.3.1.2.
This table will map each port to a corresponding port number in INTEGER format.

3. Retrieve Interface ID Table, OID for this table is .1.3.6.1.2.1.17.1.4.1.2
This table will map each port number to the port interface name.

A mapping of the port number, port name and list of the MAC Addresses of the connected devices is formulated by joining the three tables described above (Figure 2). This mapping, the Switch Look-up Table, is monitored for changes over iterations. Using this approach, it is possible to determine the devices that have been added to or removed from the network [8, 9].
Devices discovered after a device gets failed will be considered as possible replacement candidates.
Once failed devices are discovered, we search for the replacement device.

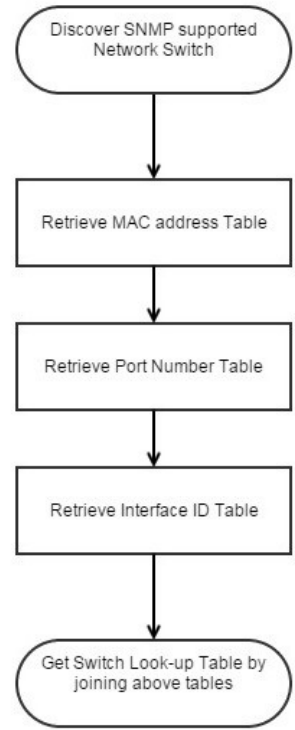

**Figure 2: Switch Port Detection**

Following are the steps to identify a replacement device (Figure 3):
1. Get the list of all failed devices. The devices which are not reachable on the audio network or being reported as failed will be considered as failed devices.
2. Get the list of available replacement devices discovered after a device failure.
3. For each failed device look for the replacement device which exists on the same port as of failed device.
4. Additionally, determine if the replacement device added on the same port matches the hardware configuration of failed device, i.e. the type of the device and its internal hardware parameters.

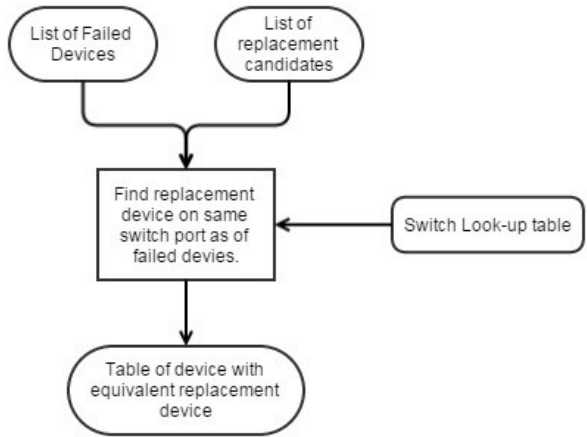

Figure 3: Identifying Replacement Devices

Once the replacement has been identified, there are three main steps towards the completion of healing (Figure 4).

1. Mapping characteristics: Characteristics of the devices include the information required to identify the device within the audio system. This is comprised of the device address (uniquely assigned by HiQnet protocol [10] for each device in a network) and the network information i.e. the IP address and DHCP settings. These settings are communicated to the replacement device, which receives the information and proceeds to modify its characteristics based on the content of the message. The system proceeds to the next stage of healing only if this step is complete.

2. Mapping firmware: In addition to hardware compatibility, it is essential that the replacement devices utilize a compatible firmware. As part of the healing procedure, correct firmware files are transferred to the device via FTP, if there is a difference in firmware version. This causes the device to update and reboot.

3. Mapping configuration: Once the characteristics and the firmware of the replacement devices have been harmonized with that of the failed device, the system proceeds to map the internal configuration.

Configuration of the device can include:
a. In case of the audio processing devices, a design/program that contains the signal processing objects, programmed logic. This program determines the real-time functioning of the audio processor.

b. The system stores the configuration of the devices in the form of XML files. These files are transferred to the appropriate replacement device via FTP. The replacement device upon receiving the files, utilizes them to run the same program as the failed device.

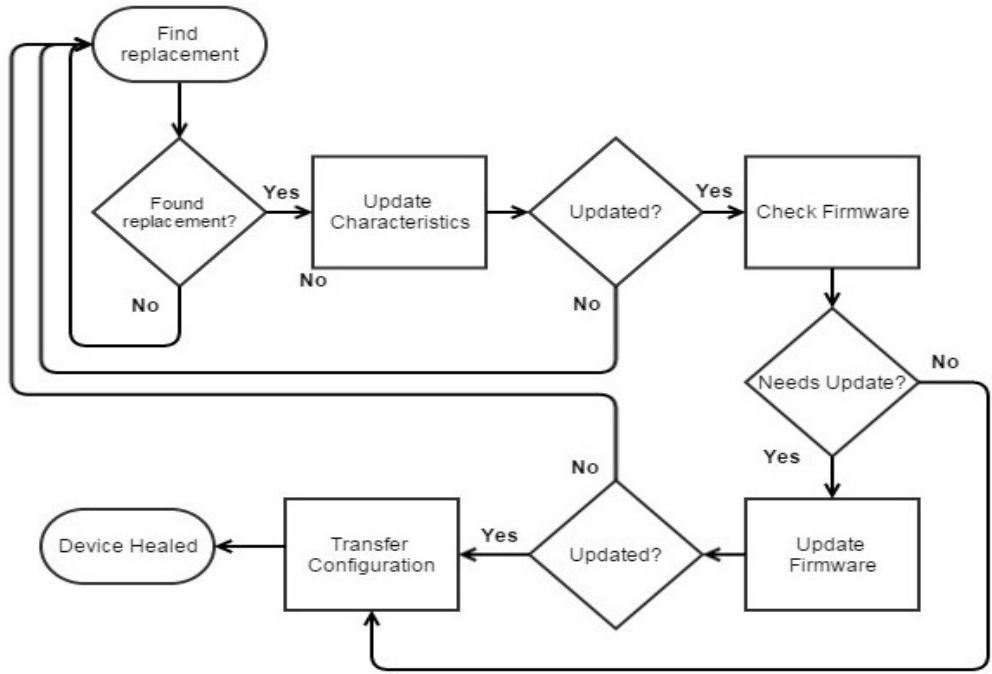

Figure 4: Configuration of the Replacement Device

## IV. EXPERIMENTAL DATA

The system proposed in this paper was implemented as part of a new Harman product called the BSS Contrio Server [11]. In this section, we present a brief overview of the experiment carried out on the product. We compare the turnover time and efficiency between manual recovery approach and recovery using our new approach.

*Experimental setup*

We assume that all devices are placed in a static network. Once deployment is done, devices are immoveable.

*A. Hardware*:

1. Crown Amplifiers [13]

2. Laptop running Harman's Control & Configuration Software Audio Architect [12]

3. BSS Contrio Server

4. HP Procurve Switch

All the above devices are connected on the same network.

*B. Software*:

Harman's Control & Configuration Software Audio Architect (AA). This application has the ability to detect Harman devices on the network and provides a way to configure and monitor the devices.

*C. Process*:

Prior to testing using both the approaches, the amplifiers were configured via AA. The number of amplifiers used was varied for each trial. In both the approaches, the failure or loss of a device was simulated by disconnecting the device from the network switch.

- Manual Recovery: After the removal of the devices from the network switch, replacement devices identical in type to the ones removed were connected to the switch. Once the replacements were detected in AA, we matched and re-mapped the configuration of the failed devices onto the new ones. At this point, the trial is deemed to have been completed. We measured the time from the point when the replacement is connected to the switch to the point when it has been configured with the settings of the failed device.

- Self-Healing: After the removal of the devices from the network switch, replacement devices identical in type to the ones removed were connected to the switch. As part of the automatic healing implemented in this approach, the system detects the loss of connectivity of the devices and also the addition of the replacement devices on the network switch. The system proceeds to 'heal' the devices and maps the last known configuration of the failed devices to the replacements. We measured the time from the point when the replacement is connected to the switch to the point when the system has finished configuring the replacement devices.

Following table displays the comparative data between naïve approach and Self-Healing system:

TABLE I. COMPARATIVE DATA

| Devices | Manual Configuration (re-configuration time taken in seconds) | Self-Healing Audio System (re-configuration time taken in seconds) |
|---|---|---|
| Crown I-Tech HD device (Quantity = 1) | 30.20 | 27.49 |
| Crown DCiN 300N device (Quantity = 1) | 32.31 | 27.78 |
| Crown I-Tech HD device (Quantity = 1), Crown DCiN 300N device (Quantity = 1) | 35.23 | 36.34 |
| Crown I-Tech HD device (Quantity = 2), Crown DCiN 300N device (Quantity = 1) | 35.41 | 36.61 |

From the results, it is apparent that the time taken by the self-healing system is comparable to the manual approach while providing the benefit of automating the whole process at the same time. Additionally, the human operator would take considerably longer time to manually map the replacement devices to the failed ones as the number of swap-outs increases. The manual approach requires that a PC/laptop is constantly connected to the network and is running the configuration software to quickly aid recovery.

## V. CONCLUSION

The method outlined in this paper for recovery of the audio system from failure attempts to improve the turnaround time for restoring the audio system to the working state.

It is worth stating that this method is built on a few assumptions which are critical to a successful recovery of the audio system.

1. The Network switch has to be managed switch because un-managed switches do not have provisions for the SNMP protocol.
2. For discovering devices connected to the network switch, the community name parameter of the network switch should be known beforehand.

As part of the future work, we intend to work on improving the device discovery method. Presently, this requires the IP address of Network Switch; it could be improved by discovering IP address of Network Switch connected to application. This method has been tested with switches from Cisco, HP and Xtreme Networks. It can be tested with more managed switches from other organizations. This can give us better knowledge about performance and the universal applicability of this approach.